# Checks and Controls in spreadsheets

Patrick O'Beirne

Systems Modelling Ltd

Tara Hill, Gorey, Co. Wexford, Ireland

Tel +353-53-942-2294 Email pob@sysmod.com

**Abstract**

*Spreadsheets that are informally created are harder to test than they should be. Simple cross-foot checks or being easily readable are modest but attainable goals for every spreadsheet developer. This paper lists some tips on building self-checking into a spreadsheet in order to provide more confidence to the reader that a spreadsheet is robust.*

*Eusprig 2009 conference*

## *Introduction*

CFO Magazine had a popular series of articles in 2008 on CFO peeves about sloppiness in spreadsheets presented to them:

http://www.cfo.com/article.cfm/11288290 Spreadsheet "Worst Practices" [1]
http://www.cfo.com/article.cfm/11525407 Sloppy Spreadsheets: Readers Speak Out [2]
http://www.cfo.com/article.cfm/11950766 Sloppier Spreadsheets: How Bad Can They Get? [3]

Many of the complaints are about spreadsheets missing cross-foot checks or not being easily readable. This paper lists some tips on building self-checking into a spreadsheet in order to provide more confidence to the reader that your spreadsheet is robust. These techniques correspond to the classical programming technique of assertions and the recent trends to test-driven development.

## *1) Cross foot*

The cross-foot check is to sum every column and every row and compare the sum of the grand total row with the sum of the row totals in the final column. Calculate the difference; if it is non-zero the amount may help you identify where the error is. Have a cell that displays a large red error indicator if the difference is significant, so that even on printouts it cannot be missed. Excel always has small rounding errors from floating point arithmetic, which although as small as 1E-13 ($10^{-13}$) will still be different from zero. Therefore you should test for the absolute value of the difference being greater than 0.01 or whatever number suits the scale of values you are working with. Here is an example of such a formula; it shows a blank if the two values are close enough, otherwise it displays a message, which for emphasis can be in formatted in a large font and red text:

=IF( ABS(H10-J10)<0.01, "", "Totals across and down do not match" )

A good sign of a careful spreadsheet maker is that the front sheet (or whatever sheet is always printed) carries forward the error check indicator from every supporting sheet, so that a report cannot be embarrassingly printed without knowing that some supporting sheet has an error.

We also want to avoid the risk of forgetting to include a row in the subtotal or total rows. One way to check the grand total is to sum the entire table range including intermediate totals, grand totals, and right-hand totals, and divide that by four.

## *2) Balance*

In Accounting 101 you learn that a balance sheet must balance; eg by defining owner's equity as a balancing line equal to total assets less total liabilities. Chemical engineers prove their calculations by a mass balance that shows how all the material inputs emerge as outputs. To prove that your spreadsheets have some data integrity, you need to look for any opportunity to verify that a total of inputs is the same as the total of outputs.

For example, if you have a total budget to allocate, the sum of the allocated amounts must equal the original total. Percentages should add up to 100%.

## *3) Proportion*

Look for opportunities to calculate averages, to divide total sales or labour costs by number of employees and compare with previous ratios. When forecasting projects, there is often no known input amount to compare with an output. In such cases we look at the absolute value of changes or the proportional value of changes to see if they correspond. One might compare last year's increase with this year's to see if they are as expected. Unfortunately, expectation is a dodgy measure that works in stable conditions but not in the unstable real world - see 'Expectations' below.

## *4) Multiple plus ungood* [1]

We've all seen this kind of total calculation:

=B11+B17+B27+B37+B48+B67

and know there is a high likelihood of a pointing error - that one of those references is pointing to the wrong cell; or that there is some reference overlooked in creating that chain. It arises when there are detail lines, intermediate totals calculated by =SUM() and to get a total of these intermediate figures, each needs to be individually selected. Here are some ways to get around this:

4a) Assuming that every detail figure is also included in the =SUM() intermediate formulas, simply calculate =SUM(B2:B67)/2.

4b) Replace all the =SUM( formulas with =SUBTOTAL(9,... Then replace the multiple-plus formula with =SUBTOTAL(9,B2:B67). The SUBTOTAL function ignores the results of other SUBTOTAL functions in its range.

If the data is an export from an accounting system or database, or is a simple table of numbers, you can get Excel to automatically put in the subtotal formulas, so you don't have to. Ensure that every column has an identifying header; and that every row has an entry in a column that indicates what group this row belongs to, and the table is sorted by that column. Then choose the Data Subtotals command, specify the grouping column, and check all the data columns that you want subtotalled; and Excel puts them in for you. When you insert or delete rows in the table, simply re-do the Subtotals command.

4c) There may be other intermediate calculations that are only required for display and not for further calculation. They could be SUM or cumulative calculations that give a running accumulation from left to right. A way to exclude their result is to use the text functions =FIXED() and =DOLLAR() which produce text results that are excluded from SUM totals. Be aware that Excel will still treat them as numbers if they are referred to individually in a formula.

---

[1] This is a play on "doubleplusungood" which in George Orwell's *1984* is Newspeak for "very bad"

## 5) Room for expansion

How do you make sure the totals still refer to the correct cells if rows are inserted or deleted? Take for example a formula in B67 =SUBTOTAL(9,B51:B66)

If you insert a row at row 51 the formula now reads =SUBTOTAL(9,B52:B67)
If you insert a row at row 67 the formula still reads =SUBTOTAL(9,B51:B66)

Either way, the row you inserted is excluded from the calculation. On the Tools > Options > Edit tab there is a setting "Extend data table formats and functions" that may automagically[2] adjust the formulas for you, but I would not rely on it.

To avoid the risk of missing an inserted row or column, always make sure every range to be summed begins and ends at a blank cell. If the first row consists only of a heading in the first column, that is readable enough. If not, you can enter in the first column the prompt "(Insert further rows below this line)" and in the last row "(Insert further rows above this line)". You could also fill the cells with ten underline characters, or make the interior colour black or blue and make the row a smaller height.

Another way to protect against insertion at the bottom of the column is: in B67 enter =SUBTOTAL(9,B51:OFFSET(B67,-1,0))

That looks like a circular reference to B67 in B67 but in fact the OFFSET function is interpreted as a reference first so in B67, OFFSET(B67,-1,0) is interpreted as B66.

http://www.mvps.org/dmcritchie/excel/offset.htm points out an alternative using INDEX to always refer to the row above:

=SUBTOTAL(9,B51:INDEX(B:B,ROW()-1))

However, both of those are more difficult for general Excel users to understand, and neither protects against insertion at the top of the range.

## 6) Other sources of information

In section (1) I gave an example of a warning message in a formula that checks cross-foot values. In programming terms this is termed an assertion. In VBA a similar concept is the Debug.Assert method which suspends execution if the test specified returns FALSE, eg

1000   Debug.Assert  ThisValue=ThatValue

In a previous Eusprig paper "Information and Data Quality in Spreadsheets" [5] I listed in Appendix 2 features of Excel such as Data Validation, Conditional Formatting, and techniques for navigating, finding, and selecting data.

The Spreadsheet Safe syllabus [6] for the Spreadsheet Safe certification examination includes many of these techniques.

---

[2] This is a play on "automatically, magically"

4                                                                    Copyright © 2009 EuSpRIG & The Author(s)

My book on Spreadsheet Check and Control [7] documents in detail 47 techniques for developing, verifying, and validating spreadsheets.

Phil Bewig's paper "How do you know your spreadsheet is right" [4] is available from the eusprig.org website.

In a paper at Eusprig 2008 [8], David Colver claimed that "Operis and smaller accountants have an integrity testing formula for every 3 to 14 formulae […], the large banks and accounting firms have such a formula for 200 to 250 formulae." He listed fourteen tests specific to the financial domain:

1. Balance sheet balances
2. Financial statements add up
3. Financial statements have expected signs
4. Sources match uses
5. Identities hold true
6. Balance sheet clears out
7. Cash cascade gives same net cash figure as the cash flow
8. Ratio Inclusion analysis [See Colver, D, Eusprig 2007 proceedings]
9. Tax reconciliation
10. Yield analysis
11. Physical identities
12. Iterative solution has converged
13. Inputs make sense
14. Outputs meet participants' requirements

## *7) Expectations*

An obvious flaw in relying on expectations is getting the answer you expect instead of the correct answer. Another is getting an answer that is materially different but not obviously so. It is unavoidable that we tend to look at the new in terms of the familiar, and when that raises questions it's useful. But when it does not raise questions, it does not mean the answer is correct - merely that that particular test gave us no information.

Remember the Dilbert cartoon where he says to a person offering a spreadsheet checking "I don't think accuracy matters if no one can tell what it's for", and that person reports back to the puzzled boss "And Dilbert found no inaccuracies!"  Some people are better than others at finding errors, just as good developers can be many times more productive than bad ones.

## 8) The Top Ten spreadsheet questions

In order to achieve quality and robustness in spreadsheet use, spreadsheet owners should provide answers to the following questions.

1. **What is the purpose of the spreadsheet?**
    a. Criticality – what if it were lost or corrupted?
2. **Where is it kept – network location, set of files**
    a. How do we know which is the current version?
    b. Complete list of data sources it depends upon
    c. What depends on this spreadsheet?
3. **How is it used? (Process documentation, instructions)**
4. **Is it for one person or is it re-used by others?**
5. **Is it once-off (project) or has it a periodic operation?**
6. **Who peer reviews its structure and version changes?**
    a. If none, likelihood of key-person risk?
    b. Evidence of test (with results) and acceptance
7. **What controls are around it?**
    a. Who reviews & signs off its output?
    b. Reconciliation with data sources
8. **What checks are included within it?**
    a. Cross-foot, balance checks, etc
9. **What evidence is there of conformity to good design practices?**
    a. Potential long list, see "Spreadsheet Check and Control" book
    b. Clear block layout, formats, print output header/footer
    c. Formula integrity, protection, no errors, no external links
    d. Use of timesaving formulas and features
10. **What are the pain points?**
    a. Quality of input data; duplication, update
    b. Grunt work transforming data
    c. Effort maintaining & updating formulas
    d. Training in more efficient Excel skills
    e. Possible to replace with controlled shared system?

© Patrick O'Beirne June 2009
www.sysmod.com